\documentclass[reprint,amsmath,amssymb,floatfix,showkeys]{revtex4-2}
\usepackage[utf8]{inputenc}
\usepackage[T1]{fontenc}
\usepackage{graphicx}
\usepackage[version=4]{mhchem}
\usepackage{booktabs}
\usepackage{multirow}
\usepackage{makecell}
\usepackage{rotating}

\begin{document}

\title{Impact of Shape Coexistence on Nuclear Stability}

\author{G.~Saxena$^{1}$}
\email{gauravphy@gmail.com}
\author{H.~Sikhwal$^{2}$}
\author{N.~Chandnani$^{1,3}$}
\author{Pranali Parab$^{4}$}
\author{Siddharth Parashari$^{5}$}
\author{Gabriela Llos\'a$^{5}$}
\author{Mamta Aggarwal$^{4}$}

\affiliation{$^{1}$Department of Physics (H\&S), Govt. Women Engineering College, Ajmer, Rajasthan-305002, India}
\affiliation{$^{2}$University of Rajasthan, Jaipur, Rajasthan-302004, India}
\affiliation{$^{3}$Department of Physics, Manipal University Jaipur, Jaipur-303007, India}
\affiliation{$^{4}$Department of Physics, University of Mumbai, Vidhyanagari, Mumbai, 400098, Maharashtra, India}
\affiliation{$^{5}$Instituto de F\'isica Corpuscular (IFIC), CSIC-UV, Valencia, Spain}

\begin{abstract}
Nuclear decay properties are conventionally predicted assuming nuclei decay from their ground-state configurations. However, this often neglects a fundamental structural complexity which is the phenomenon of shape coexistence, where nuclei possess multiple competing configurations at nearly degenerate energies. When both parent and daughter nuclei can exist in different energy minima, multiple decay pathways become possible. We systematically investigate how shape coexistence influences nuclear decay for approximately 1500 even-even nuclei ($8 \leq Z \leq 118$, $8 \leq N \leq 184$) using the Nilsson-Strutinsky method and relativistic mean-field calculations with NL3$^*$, DD-ME2, and DD-PC1 functionals. We identify around 400 nuclei exhibiting competing energy minima separated by less than 1 MeV. For these shape-coexisting nuclei, we calculate $\alpha$, $\beta^+$ and $\beta^-$ decay half-lives considering all possible transition pathways between the competing minima. Our results demonstrate that shape coexistence substantially impacts decay predictions, with half-lives showing variations up to nearly one logarithmic unit depending on which configurations participate in the transition. Comparison with experimental data from NUBASE2020 shows that pathways involving the second minimum sometimes reproduce measured lifetimes more closely than conventional ground-state to ground-state assumptions. Branching ratios exhibit even stronger sensitivity, with certain nuclei displaying complete inversions of the dominant decay mode depending on configuration choice. These pathway-dependent variations are not due to model uncertainties but reflect inherent structural effects. The correlation between the shape dynamics and nuclear stability establishes the shape coexistence as an essential component in predictive nuclear structure and astrophysics studies.
\end{abstract}


\maketitle

\section{Introduction}

Precise measurements and the knowledge of various properties of nuclei including nuclear masses, binding energies, decay mechanism and half-lives of nuclei that lie beyond the line of stability serve as crucial input for astrophysical investigations aimed at understanding energy production in stars, supernova explosions and nucleosynthesis r-process~\cite{cowan2021origin, thielemann2017neutron, mumpower2016impact}. While modern radioactive ion beam facilities have extended our experimental reach into neutron-rich territory, the vast majority of nuclei participating in the r-process remain experimentally inaccessible, making reliable theoretical predictions indispensable. These theoretical predictions should  not only reproduce the known experimental systematics but also extrapolate to unmapped regions of the nuclear chart where direct measurements are not feasible~\cite{erler2012limits,marketin2016large}.

Deformed nuclei across all mass regions are known to exhibit characteristic rotational spectra, electromagnetic transition rates and modified single-particle schemes that alter the decay properties~\cite{casten2000nuclear, raman2001transition, Coban2012, Misicu1998}. Despite its recognized importance, the explicit role of deformation in modifying decay rates, particularly in regions where nuclei can have multiple competing shapes, remains inadequately incorporated into predictive decay models~\cite{Nabi2024, Raduta1993, Chen2023, jain2024shape_scientific_reports}.

In certain mass regions, competition between shell effects, pairing correlations and macroscopic surface energy leads to shape coexistence, where two or more distinct nuclear configurations correspond to local minima in the potential energy surface (PES) separated by small energy differences \cite{heyde2011shape, garrett2024beta, Leoni2024_PPNP, Leoni2024_EPJST, Garrett2022, Yang2023}. There has been ample experimental evidence, established through electromagnetic observables, laser spectroscopy and transfer reactions, confirming shape coexistence in regions including the Pb-Hg isotopes around $Z \approx 82$ where $^{186}$Pb exhibits triple shape coexistence~\cite{Andreyev2000, Helariutta1999}, the $A \approx 100$ region at $N=60$ (e.g., $^{98}$Sr, $^{100}$Zr)~\cite{Wu2004, Clement2016}, the Cd-Sn region ($^{110,112}$Cd)~\cite{Garrett2019}, and light nuclei around the $N=20$ island of inversion~\cite{32Mg, 34Si, N20}. Theoretical predictions extend more broadly. Mean-field approaches, including Hartree-Fock-Bogoliubov with Skyrme and Gogny interactions~\cite{Bender2003, Decharge1980}, the Nilsson-Strutinsky method~\cite{Strutinsky1967, BrackDamgaard1972}, and relativistic mean-field calculations~\cite{lalazissis2009effective, lalazissis2005new, nikvsic2008relativistic}, identify competing configurations across extensive regions. The Finite Range Droplet Model~\cite{moller2016nuclear}, Interacting Boson Model, and shell model calculations~\cite{Iachello2006, Caurier2005} provide complementary frameworks. Beyond confirmed cases, these models also predict shape coexistence in the $N \approx 28$-40 transitional zone~\cite{Hamilton1974}, neutron-rich $N \approx 90$ rare-earths~\cite{Kulp2008, Polikanov1962}, and actinides near $N=126$~\cite{Girod1989}, suggesting widespread occurrence across the nuclear chart.

Shape coexistence impacts nuclear stability by altering binding energies through the competition between shell gaps and deformation-driving interactions, leading to shell quenching near magic numbers and the emergence of shape isomers with measurable lifetimes as seen in our recent works \cite{AGGARWAL2024122843, AGGARWAL2026140298}. Decay properties become particularly sensitive as $\beta$-decay half-lives depend on shell structure while Q-values and barrier penetration vary with configuration choice, substantially affecting predicted lifetimes and branching ratios~\cite{heyde2011shape, mumpower2016impact}.

Despite extensive investigation~\cite{heyde2011shape, garrett2024beta}, explicit incorporation of shape coexistence into decay predictions remains limited. Phenomenological formulas widely used for half-life predictions are parameterized in terms of mass, charge and Q-value~\cite{saxena2021new, cheng2022isospin, royer2020, ismail2022improved} but do not account for deformation or distinguish between configurations. Microscopic approaches based on density functional theory~\cite{GORIELY199628} or QRPA~\cite{marketin2016large} can capture deformation effects but are computationally restricted. The absence of a deformation-dependent, efficient framework addressing shape coexistence represents a significant gap in predicting decay properties reliably in transitional regions.

This gap is particularly consequential for nucleosynthesis studies, where astrophysical network calculations modeling the r-process, rp-process, and p-process are critically sensitive to half-lives and branching ratios of key waiting-point nuclei~\cite{schatz1998rp, arcones2011review, mumpower2016impact, surman2014sensitivity}. Uncertainties in half-life predictions propagate directly into uncertainties in abundance patterns, freeze-out timescales and heavy element synthesis.

In this work, we systematically investigate how shape coexistence influences decay dynamics across the nuclear chart. We map potential-energy surfaces for even-even nuclei using the Nilsson-Strutinsky Method (NSM) and relativistic mean-field calculations with the non-linear meson-nucleon coupling (NL3$^*$), the density-dependent meson-exchange (DD-ME2) and the density-dependent point-coupling (DD-PC1) models. We identify nuclei with competing, near-degenerate minima and analyze how transitions between different configurations in both parent and daughter nuclei affect Q-values, half-lives and branching ratios. Calculating $\alpha$, $\beta^+$ and $\beta^-$-decay lifetimes while incorporating deformation and shape effects along with coexisting states, we compare predictions with experimental data from NUBASE2020 to quantify the role of deformation and shape coexistence in determining nuclear decay properties.

\section{Details of Calculations}

The theoretical framework employed in this study utilizes a complementary dual-model approach, combining the Nilsson-Strutinsky Method (NSM) \cite{Strutinsky1967, BrackDamgaard1972} with Relativistic Mean-Field (RMF) \cite{ring1996relativistic} calculations. This selection is strategically designed to provide a robust cross-comparison between the macroscopic-microscopic framework, which serves as the physical foundation for global benchmarks like the Finite Range Droplet Model (FRDM) and modern self-consistent covariant density functional theory. To ensure that the identified shape coexistence is a genuine physical feature rather than an artifact of a specific model's parametrization, we incorporate three distinct RMF functionals: NL3$^*$ \cite{lalazissis2009effective}, DD-ME2 \cite{lalazissis2005new} and DD-PC1 \cite{nikvsic2008relativistic} models. The numerical details of these calculations, including basis size convergence tests, mathematical formulations of the pairing interactions, and implementation specifics of these theoretical models, are provided in the Supplementary Material.

A comprehensive set of calculations has been performed employing the above mentioned models to investigate the variation of binding energy as a function of the quadrupole deformation parameter ($\beta$) for all even-even nuclei in the range: 8$\leq$Z$\leq$118 and 8$\leq$N$\leq$184. In total, approximately 1,500 even–even nuclei have been analyzed. For each nucleus, the corresponding potential energy surface (PES) has been computed and plotted with respect to its quadrupole deformation. This approach has enabled the identification of the ground-state shapes and the study of shape coexistence, as well as transitions between spherical, prolate, and oblate configurations across the nuclear chart. To ensure computational feasibility, this analysis is restricted to axially symmetric prolate and oblate shapes, a standard convention in mean-field Potential Energy Surface (PES) calculations \cite{GROSSE2022137328, PEREZ2017363}. Although triaxiality influences the dynamics of transitional nuclei (e.g., $A \approx 70-100$) \cite{esmaylzadeh2022investigation}, its inclusion increases computational feasibility \cite{ryssens2016} without fundamentally altering the qualitative identification of near-degenerate minima, as ground-state masses are often only weakly sensitive to axial symmetry breaking in these regions \cite{GROSSE2022137328}.

\begin{figure}[ht]
\centering
\includegraphics[width=\linewidth]{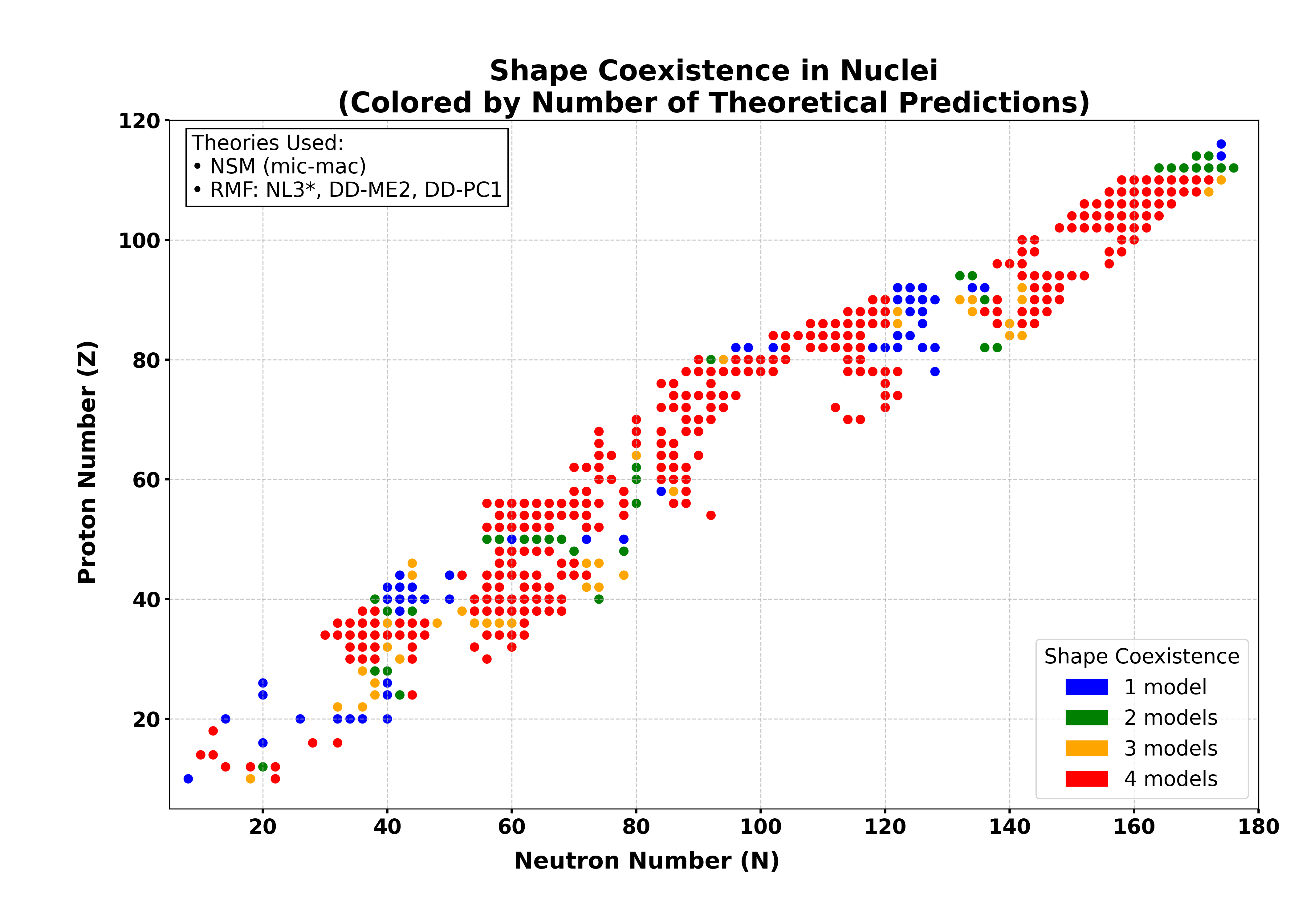}
\caption{N-Z chart displaying nuclei exhibiting at least two distinct minima in their potential energy surfaces (PES) as predicted by NSM, NL3$^*$, DD-ME2, and DD-PC1. Color coding indicates the number of theoretical models predicting shape coexistence for each nucleus: red markers represent nuclei where all four models agree, blue markers indicate three models, green indicates two models, and orange represents predictions from a single model.}
\label{n-z-chart-400}
\end{figure}

Our calculations across all four theoretical frameworks (NSM, NL3*, DD-ME2, and DD-PC1) identify approximately 400 nuclei that exhibit shape coexistence. The selection is based on a specific energetic criterion: the presence of at least two distinct minima in the PES separated by less than 1 MeV. These nuclei are presented in Fig.~\ref{n-z-chart-400} as an N–Z chart. A significant outcome of this analysis is the number of nuclei that demonstrate consistent shape coexistence across all four models, as indicated by the dominance of red-colored markers in the figure. The islands of shape coexistence appear as distinct bands spanning diverse regions of the nuclear chart. A prominent cluster is visible in the light-to-medium mass region around $Z \sim 38$--$40$ and $N \sim 58$--$62$, corresponding to the well-documented Zr--Sr region where nuclei such as $^{98,100}$Sr and $^{100,102}$Zr exhibit competing configurations confirmed by spectroscopic studies~\cite{heyde2011shape, Wood1992, Clement2016, Wu2004}. Another dense band appears near $Z \sim 48$--$50$ and $N \sim 60$--$70$, the Cd--Sn region where $^{110,112}$Cd exhibit multiple shape coexistence~\cite{Garrett2019}. A horizontal band along $Z \sim 82$ between $N \sim 104$--$110$, representing the Pb--Hg region where $^{186,188}$Pb and neutron-deficient Hg isotopes display shape coexistence~\cite{Andreyev2000, Helariutta1999}. Additional clustering is evident in the rare-earth region ($Z \sim 62$--$74$, $N \sim 88$--$104$) where nuclei such as $^{152}$Sm have been confirmed through electromagnetic measurements~\cite{Kulp2008}, and scattered points extend into the superheavy region ($Z \geq 100$). In most cases, the identified candidates demonstrate agreement with Finite Range Droplet Model (FRDM) predictions~\cite{moller2016nuclear}, establishing a degree of consistency between macroscopic-microscopic and self-consistent mean-field frameworks. This alignment suggests that shape coexistence is a frequent structural phenomenon across various regions of the nuclear chart. Although multiple shapes may coexist for a given nucleus, the dominant configuration, defined as the minimum with the lowest binding energy, does not always align across models. This variation underscores the inherent model dependence in predicting the ground-state shape of a nucleus.

\begin{figure}[ht]
\centering
\includegraphics[width=\linewidth]{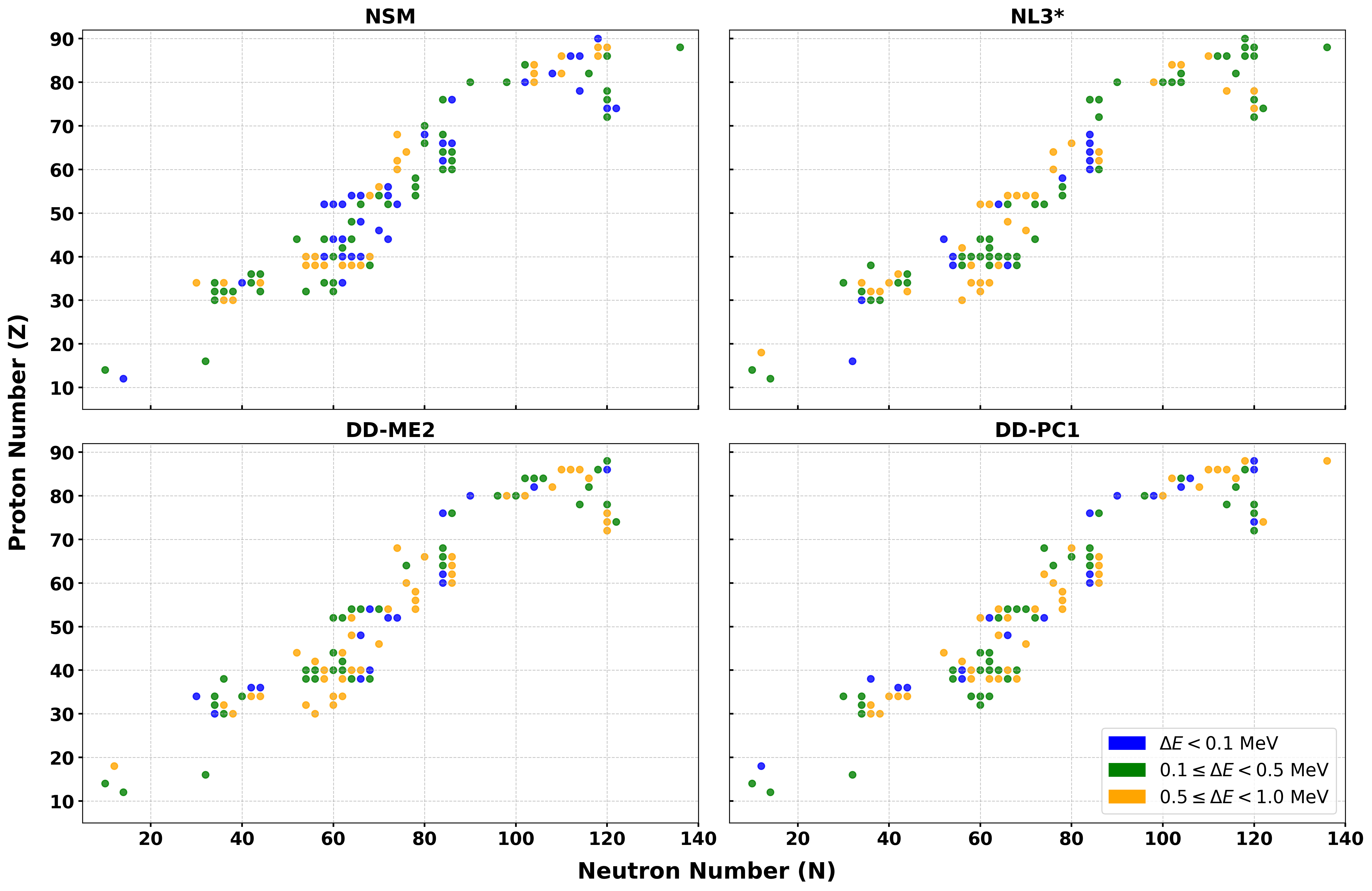}
\caption{N--Z chart for the refined set of 80 nuclei from Fig.~\ref{n-z-chart-400} exhibiting robust shape coexistence within the energy threshold $\Delta E \leq 1.0$ MeV across all four theoretical models (NSM, NL3$^*$, DD-ME2, DD-PC1). Each panel displays results from one model. Color coding indicates the energy separation between competing minima: blue markers represent extremely small separations ($\Delta E < 0.1$ MeV), green markers indicates separations $0.1 \leq \Delta E < 0.5$ MeV and orange markers indicate moderate separations ($0.5 \leq \Delta E < 1.0$ MeV).}
\label{n-z-chart-121}
\end{figure}

\begin{figure*}[ht]
  \centering
  \includegraphics[width=\linewidth]{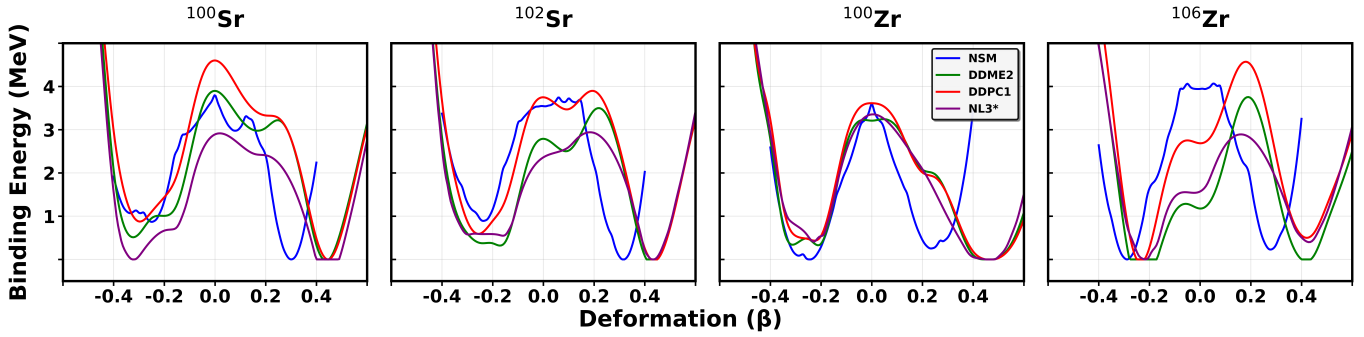}
  \caption{Potential energy surfaces (PES) as functions of quadrupole deformation ($\beta$) for $^{100,102}$Sr and $^{100,106}$Zr, calculated with NSM, NL3$^*$, DD-ME2, and DD-PC1. The presence of multiple competing minima separated by small energy differences confirms shape coexistence in these nuclei. The different colors represent predictions from the four theoretical models, illustrating both inter-model consistency and variation in the detailed structure of potential energy landscapes.}
  \label{Fig:shape-coexistence}
  \end{figure*}

While Fig.~\ref{n-z-chart-400} identifies the global distribution of shape coexisting nuclei, not all cases are equally relevant for studying observable effects on decay properties. The nuclei presented exhibit varying energy separations between their competing minima, ranging from a few keV to several MeV depending on the specific nucleus and theoretical model. Shape coexistence effects become most pronounced when the competing configurations are energetically comparable, making such nuclei the most likely candidates to exhibit shape transitions and enhanced collective excitations. To focus on these cases, we impose a stringent selection criterion: any nucleus with an energy separation exceeding 1.0 MeV in any of the models is excluded. Applying this criterion yields a refined set of 80 nuclei exhibiting shape coexistence (Fig.~\ref{n-z-chart-121}), for which all models confirm near-degeneracy within $\Delta E \leq 1.0$ MeV. This refined group generally aligns with the energetic trends predicted by FRDM calculations~\cite{moller2016nuclear}, providing a consistent cross-reference for their identification. Such small energy separations allow transitions between competing configurations with minimal energetic input, enabling enhanced sensitivity to external perturbations (e.g., temperature, angular momentum)~\cite{AGGARWAL2026140298} and facilitating quantum tunneling or mixing between prolate and oblate states. To ensure physical relevance, we further require that the identity of the dominant minimum be consistent across all four models. Cases where models disagree on the lowest-energy shape are excluded from further analysis.

From this refined set, we identified 24 nuclei for which experimental half-lives and decay modes are available in NUBASE2020~\cite{wang2021ame}. We select $^{100,102}$Sr and $^{100,106}$Zr as representative cases (Fig.~\ref{Fig:shape-coexistence}) as they belong to the $A \approx 100$ mass region, a well-established benchmark where rapid structural transitions and shape coexistence are experimentally documented~\cite{Clement2016, Wu2004}. This choice allows for a transparent validation of the theoretical results against known physical phenomena. For consistency across models, binding energies are normalized to zero at each framework's global minimum. The characteristic double-well structures are clearly visible for these isotopes across all four models. In $^{100,102}$Sr, the frameworks consistently predict nearly degenerate minima at $\beta \approx \pm 0.2$ with energy separations typically found between $\Delta E \approx 0.2$-$0.5$ MeV. Similarly, for $^{100,106}$Zr, all models exhibit pronounced double-well structures, specifically with minima near $\beta \approx -0.2$ and $\beta \approx +0.4$, while maintaining a narrow energy proximity of $\Delta E \approx 0.3$-$0.8$ MeV. These small energy separations are physically significant, this indicates that nuclei can exist in competing configurations, which creates multiple possible decay pathways. When calculating decay half-lives, the choice of which minimum represents the parent or daughter state affects the Q-value, and consequently the predicted lifetime.

\section{Results}

For these identified candidates, the experimental decay characteristics documented in NUBASE2020 \cite{wang2021ame} exhibit several features.
\begin{enumerate}
    \item Eighteen (18) nuclei decay through a single dominant mode with 100\% decay probability. Among these, one nucleus undergoes pure $\alpha$-decay, six nuclei decay via $\beta^+$/EC-decay, and the remaining eleven nuclei decay through $\beta^-$-decay.
    \item Six (6) nuclei exhibit two competing decay modes, consisting of $\alpha$-decay and $\beta^+$/EC-decay.
\end{enumerate}

This decay diversity among the shape-coexisting candidates highlights their structural richness and provides a fertile ground for exploring  the potential influence of nuclear shape on decay lifetime. To quantify this influence, we developed deformation-dependent semi-empirical formulas for the three decay modes, augmenting standard phenomenological  expressions with an explicit deformation term derived from Bayesian Model Averaging of four nuclear structure models (FRDM, HFB, WS4, RMF). The complete mathematical formulation, fitting procedure, performance metrics, and structured extrapolation tests validating these models in unexplored nuclear regions are detailed in the Supplementary Material.

A crucial aspect of the present analysis is the examination of how decay properties change when different intrinsic configurations of parent and daughter nuclei are involved. For the nuclei in our final selection, all four theoretical models consistently predict that the primary and secondary PES minima are separated by an energy difference of less than 1.0 MeV. This requirement ensures that we focus only on regions where shape competition is robust and the configurations are nearly degenerate. All four pathways below describe ground-state to ground-state decay transitions, differing only in which energetically competing configuration is assumed for each nucleus:
\begin{itemize}\item \textbf{Ground-to-Ground (G--G$'$)}: both parent and daughter are assigned their global PES minima.
\item \textbf{Ground-to-Second (G--S$'$)}: the parent occupies its global minimum; the daughter is assigned its secondary minimum.
\item \textbf{Second-to-Ground (S--G$'$)}: the parent is assigned its secondary minimum; the daughter occupies its global minimum.
\item \textbf{Second-to-Second (S--S$'$)}: both parent and daughter are assigned their respective secondary minima.
\end{itemize}

The transition labels G–G$'$, G–S$'$, S–G$'$, and S–S$'$ should not be interpreted as spectroscopic decay branches between experimentally observed excited states, nor do they involve calculations of electromagnetic transition probabilities. In nuclei exhibiting shape coexistence, electromagnetic decay from the secondary minimum to the ground-state configuration of the same nucleus may occur with significant probability. However, this does not exclude the possibility of decay proceeding directly from the secondary minimum to states of the daughter nucleus. In the present work, these transition labels are therefore used to represent alternative intrinsic configurations associated with different minima of the potential-energy surfaces of the parent and daughter nuclei. By considering all such configurations, we examine how decay observables depend on the underlying nuclear structure when competing shapes are present. The analysis thus focuses on the static structural impact of shape coexistence on decay energies ($Q$-values) and the resulting sensitivity of predicted decay properties to intrinsic shape configurations.

\begin{figure}[ht]
\centering
\includegraphics[width=\linewidth]{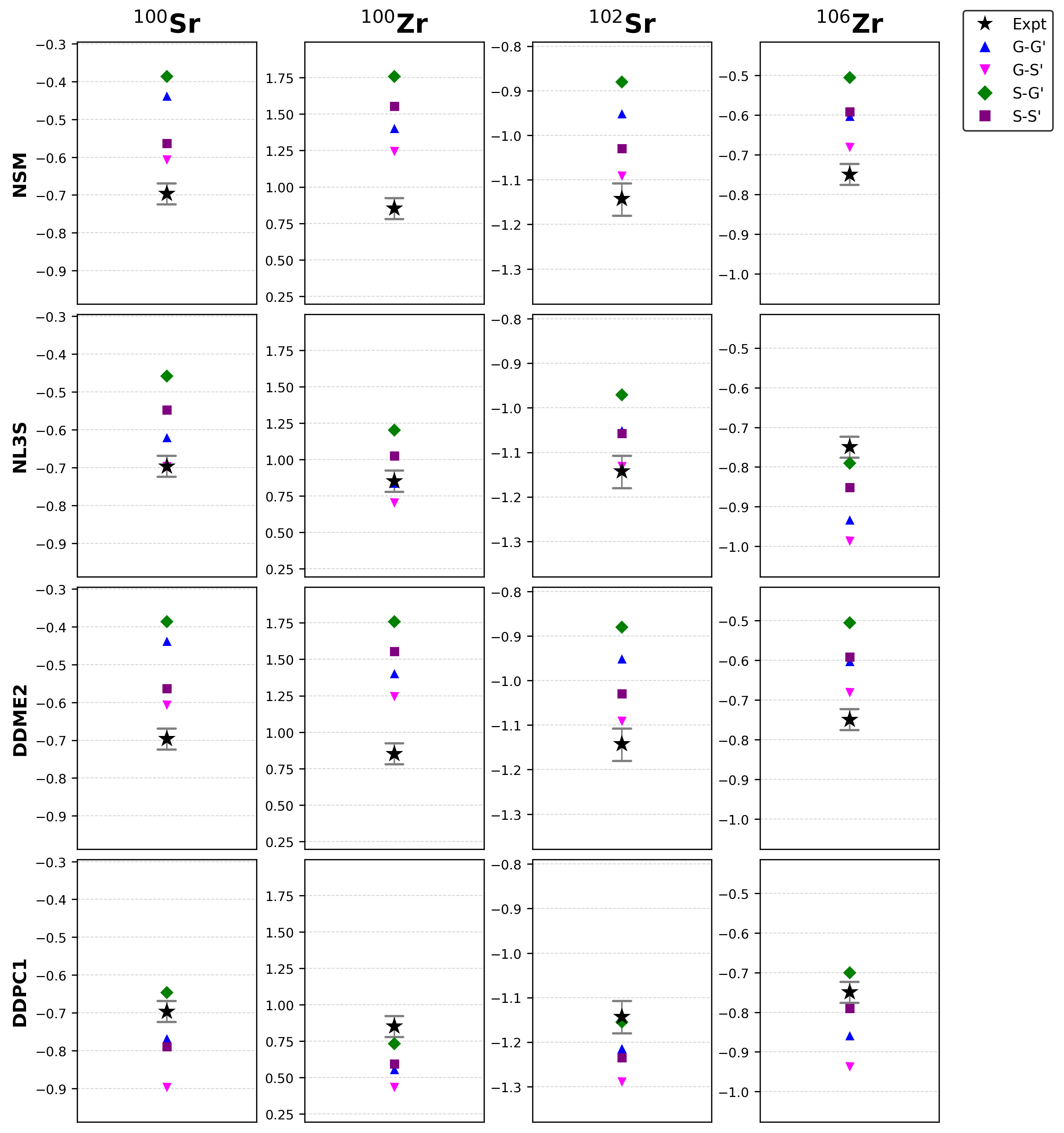}
\caption{Predicted $\log_{10}T_{1/2}$ values (in seconds) for shape-coexisting nuclei $^{100,102}$Sr and $^{100,106}$Zr using deformation-dependent formulas across different transition pathways. Each panel corresponds to one nucleus, with rows representing different theoretical models (NSM, NL3$^*$, DD-ME2, DD-PC1). The four colored points in each panel represent the four possible transition pathways: ground-to-ground (G--G$'$), ground-to-second minimum (G--S$'$), second-to-ground (S--G$'$), and second-to-second minimum (S--S$'$). The horizontal black line representing the experimental uncertainty.}
\label{fig:halflife}
\end{figure}

Fig.~\ref{fig:halflife} shows calculated half-lives for representative shape-coexisting nuclei ($^{100,102}$Sr and $^{100,106}$Zr) across all four models and transition pathways. Each panel displays results from one theoretical model (NSM, NL3$^*$, DD-ME2, DD-PC1), with the four colored points representing the four transition pathways (G--G$'$, G--S$'$, S--G$'$, S--S$'$) for each nucleus. The results demonstrate a fundamental impact of shape coexistence: when both parent and daughter nuclei can exist in multiple energy minima, the choice of transition pathway substantially alters predicted lifetimes. Depending on which energy minimum the parent and daughter nuclei occupy, predicted half-lives can vary by 0.3 to 0.9s logarithmic units for the same nucleus. For $^{100}$Sr (experimental value: $\log_{10}T_{1/2} \approx -0.70$s), the four pathways span predictions from approximately $-0.85$s to $-0.40$s across different models. In $^{100}$Zr (experimental: $\approx 0.88$s), predictions range from $\sim$0.7s to 1.75s depending on the pathway. Similar spreads appear in $^{102}$Sr and $^{106}$Zr. This variation is not due to model uncertainty, all four pathways use the same theoretical framework, but rather reflects the structural multiplicity inherent in shape coexistence. The pathway providing closest agreement with experiment differs from nucleus to nucleus. For some cases like $^{100}$Sr, the G--S$'$ pathway matches best; for others like $^{102}$Sr, the S--S$'$ or G--S$'$ combinations perform better; while $^{100}$Zr favors G-G$'$. This demonstrates that shape coexistence is not a theoretical artifact but a physical reality: the actual decay process depends on which configurations the parent and daughter nuclei occupy, and this varies across the nuclear chart. Accurate lifetime predictions in transitional regions require explicit consideration of competing configurations rather than conventional ground-state to ground-state assumptions.

The half-life analysis above provides only a partial view since several channels can compete simultaneously. Branching ratios offer a more stringent probe, reflecting the relative contributions of different decay paths. The branching ratio for decay mode $i$ is defined as
\begin{equation}
BR_i = \frac{T^{\mathrm{Th}}_{1/2}}{T^{i}_{1/2}}, \quad
i \in \{\alpha,\,\beta^{-},\,\beta^{+}/\mathrm{EC}\},
\label{eq:branching}
\end{equation}
where the total theoretical half-life satisfies
\begin{equation}
\frac{1}{T^{\mathrm{Th}}_{1/2}} = \frac{1}{T^{\alpha}_{1/2}}+ \frac{1}{T^{\beta^-}_{1/2}} + \frac{1}{T^{\beta^+/\mathrm{EC}}_{1/2}}.
\label{eq:total}
\end{equation}

\begin{sidewaystable}
\centering
\caption{Calculated branching ratios for selected nuclei across different theoretical models (NSM, NL3$^*$, DD-ME2, DD-PC1) and shape-transition scenarios (G--G$'$, G--S$'$, S--G$'$, S--S$'$). Experimental reference branching ratios are listed alongside for comparison. Dashes (--) indicate energetically forbidden transitions.}
\label{tab:branching}
\renewcommand{\arraystretch}{1.2}
\setlength{\tabcolsep}{1.8pt}
\scriptsize
\begin{tabular}{|l|l||rr|rr|rr|rr||rr|rr|rr|rr||rr|rr|rr|rr||rr|rr|rr|rr|}
\hline
\multirow{3}{*}{Nucleus} &
\multirow{3}{*}{\makecell{Exp.\\Mode}} &
\multicolumn{8}{c||}{NSM} &
\multicolumn{8}{c||}{NL3$^*$} &
\multicolumn{8}{c||}{DD-ME2} &
\multicolumn{8}{c|}{DD-PC1} \\
\cline{3-34}
& &
\multicolumn{2}{c|}{GG} & \multicolumn{2}{c|}{GS} & \multicolumn{2}{c|}{SG} & \multicolumn{2}{c||}{SS} &
\multicolumn{2}{c|}{GG} & \multicolumn{2}{c|}{GS} & \multicolumn{2}{c|}{SG} & \multicolumn{2}{c||}{SS} &
\multicolumn{2}{c|}{GG} & \multicolumn{2}{c|}{GS} & \multicolumn{2}{c|}{SG} & \multicolumn{2}{c||}{SS} &
\multicolumn{2}{c|}{GG} & \multicolumn{2}{c|}{GS} & \multicolumn{2}{c|}{SG} & \multicolumn{2}{c|}{SS} \\
\cline{3-34}
& &
$\alpha$ & $\beta$ & $\alpha$ & $\beta$ & $\alpha$ & $\beta$ & $\alpha$ & $\beta$ &
$\alpha$ & $\beta$ & $\alpha$ & $\beta$ & $\alpha$ & $\beta$ & $\alpha$ & $\beta$ &
$\alpha$ & $\beta$ & $\alpha$ & $\beta$ & $\alpha$ & $\beta$ & $\alpha$ & $\beta$ &
$\alpha$ & $\beta$ & $\alpha$ & $\beta$ & $\alpha$ & $\beta$ & $\alpha$ & $\beta$ \\
\hline\hline

$^{152}$Er &
\makecell{$\alpha$: 90\%\\$\beta^+$/EC: 10\%} &
0.99 & 0.01 & 1.00 & 0.00 & 0.91 & 0.09 & 0.99 & 0.01 &
1.00 & 0.00 & -- & -- & 0.99 & 0.01 & -- & -- &
1.00 & 0.00 & -- & -- & 0.99 & 0.01 & -- & -- &
1.00 & 0.00 & -- & -- & 0.99 & 0.01 & -- & -- \\
\hline

$^{178}$Hg &
\makecell{$\alpha$:89\%\\$\beta^+$/EC: 11\%} &
1.00 & 0.00 & 0.00 & 1.00 & 1.00 & 0.00 & 0.00 & 1.00 &
0.00 & 1.00 & 0.00 & 1.00 & 0.01 & 0.99 & 0.01 & 0.99 &
0.00 & 1.00 & 0.00 & 1.00 & 0.24 & 0.76 & 0.00 & 1.00 &
0.47 & 0.53 & 0.00 & 1.00 & 0.57 & 0.43 & 0.00 & 1.00 \\
\hline

$^{186}$Pb &
\makecell{$\alpha$:40\%\\$\beta^+$/EC: 60\%} &
0.88 & 0.12 & 0.21 & 0.79 & 1.00 & 0.00 & 0.97 & 0.03 &
0.00 & 1.00 & 0.00 & 1.00 & 0.00 & 1.00 & 0.00 & 1.00 &
0.12 & 0.88 & 0.00 & 1.00 & 0.22 & 0.78 & 0.00 & 1.00 &
0.66 & 0.34 & 0.00 & 1.00 & 0.79 & 0.21 & 0.00 & 1.00 \\
\hline

$^{198}$Rn &
\makecell{$\alpha$:90\%\\$\beta^+$/EC: 10\%} &
0.99 & 0.01 & 0.98 & 0.02 & 1.00 & 0.00 & 0.99 & 0.01 &
0.60 & 0.40 & 0.00 & 1.00 & 0.96 & 0.04 & 0.00 & 1.00 &
0.63 & 0.37 & 1.00 & 0.00 & 0.00 & 1.00 & 0.14 & 0.86 &
0.91 & 0.09 & 0.00 & 1.00 & 1.00 & 0.00 & 0.44 & 0.56 \\
\hline

$^{206}$Rn &
\makecell{$\alpha$:62\%\\$\beta^+$/EC: 38\%} &
0.90 & 0.10 & 0.10 & 0.90 & 0.96 & 0.04 & 0.28 & 0.72 &
0.00 & 1.00 & -- & -- & 0.00 & 1.00 & -- & -- &
0.00 & 1.00 & 0.00 & 1.00 & 0.00 & 1.00 & 0.00 & 1.00 &
0.00 & 1.00 & 0.00 & 1.00 & 0.00 & 1.00 & 0.00 & 1.00 \\
\hline

$^{208}$Ra &
\makecell{$\alpha$:87\%\\$\beta^+$/EC: 13\%} &
0.63 & 0.37 & 0.00 & 1.00 & 0.99 & 0.01 & 0.21 & 0.79 &
0.00 & 1.00 & -- & -- & 0.00 & 1.00 & -- & -- &
0.00 & 1.00 & 0.00 & 1.00 & 0.00 & 1.00 & 0.00 & 1.00 &
0.00 & 1.00 & 0.00 & 1.00 & 0.00 & 1.00 & 0.00 & 1.00 \\
\hline
\end{tabular}
\end{sidewaystable}

Table~\ref{tab:branching} shows the impact of shape coexistence in decay probabilities. A value of 0.00 indicates a finite but extremely small probability; a dash (-) indicates an energetically forbidden mode. We focus on cases exhibiting substantial pathway-dependent variations. In $^{178}$Hg (expt.: $\alpha = 89\%$, $\beta^+$/EC $= 11\%$), the pathway selection determines the dominant decay mode. NSM predicts 100\% $\alpha$ dominance through G-G$'$ and S-G$'$, but complete $\beta^+$/EC dominance (100\%) through G-S$'$ and S-S$'$. DD-PC1 shows similar sensitivity: G-G$'$ yields 47\% $\alpha$ and S-G$'$ gives 57\% $\alpha$, while G-S$'$ and S-S$'$ both produce 100\% $\beta^+$/EC. This indicates that the daughter nucleus configuration controls which decay channel dominates. Significant effects also appear in $^{186}$Pb (expt.: $\alpha = 40\%$, $\beta^+$/EC $= 60\%$). NSM predicts 88\% $\alpha$ branching through G-G$'$ and 100\% through S-G$'$, but G-S$'$ inverts to 79\% $\beta^+$/EC. DD-PC1 shows 66\% $\alpha$ (G-G$'$) and 79\% $\alpha$ (S-G$'$), yet G-S$'$ and S-S$'$ both yield 100\% $\beta^+$/EC dominance. In $^{198}$Rn (expt.: $\alpha = 90\%$, $\beta^+$/EC $= 10\%$), DD-ME2 exhibits complete mode reversal depending on pathway: G-S$'$ produces 100\% $\alpha$ while S-G$'$ produces 100\% $\beta^+$/EC. NL3$^*$ and DD-PC1 show similar inversions between specific pathways. For $^{206}$Rn and $^{208}$Ra, NSM demonstrates pathway dependence with branching ratios varying from $\alpha$-dominant (90\% and 63\% through G-G$'$) to $\beta^+$/EC-dominant (90\% and 100\% through G-S$'$). The RMF models predict consistent $\beta^+$/EC dominance across all pathways for these nuclei. In contrast, $^{152}$Er shows relatively stable branching ratios, with $\alpha$ dominance (91-100\%) maintained across most accessible pathways, indicating that pathway sensitivity varies across different nuclei. These results establish that the branching ratios in shape-coexisting systems are configuration-dependent. The complete inversions of dominant decay modes across different pathways, including cases where the same nucleus switches between 100\% $\alpha$-decay and 100\% $\beta^+$/EC-decay, demonstrate that nuclear shape influences both absolute decay rates and the competition between decay channels.
\begin{table*}[tp]
  \centering
  \caption{Predicted $\log_{10}T_{1/2}$ values (in seconds) for selected nuclei using 
  deformation-dependent semi-empirical formulas. Results shown for different transition 
  pathways: ground-to-ground (G-G'), ground-to-second minima (G-S'), second 
  minima-to-ground (S-G') and second minima-to-second minima (S-S').}
  \label{tab:logt_half}
  \small
  \resizebox{\textwidth}{!}{%
  \begin{tabular}{llrrrr|rrrr|rrrr|rrrr}
  \toprule
  \multirow{2}{*}{Nucleus} & 
  \multirow{2}{*}{\makecell{Predicted\\Decay Mode}}&
  \multicolumn{4}{c}{NSM} &
  \multicolumn{4}{c}{NL3$^*$} &
  \multicolumn{4}{c}{DDME2} &
  \multicolumn{4}{c}{DDPC1} \\
  \cmidrule(lr){3-6}\cmidrule(lr){7-10}\cmidrule(lr){11-14}\cmidrule(lr){15-18}
  & & G-G$'$ & G-S$'$ & S-G$'$ & S-S$'$ &
      G-G$'$ & G-S$'$ & S-G$'$ & S-S$'$ &
      G-G$'$ & G-S$'$ & S-G$'$ & S-S$'$ &
      G-G$'$ & G-S$'$ & S-G$'$ & S-S$'$ \\
  \midrule
  $^{92}$Se   & $\beta^-$ & $-0.9933$ & $-0.9304$ & $-1.0205$ & $-0.9593$ & $-0.9740$ & $-0.8325$ & $-1.1044$ & $-0.9806$ & $-1.1756$ & $-1.0685$ & $-1.2911$ & $-1.1962$ & $-1.1427$ & $-1.0860$ & $-1.1936$ & $-1.1398$ \\
  $^{94}$Se   & $\beta^-$ & $-1.4631$ & $-1.4201$ & $-1.5131$ & $-1.4722$ & $-1.5443$ & $-1.2913$ & $-1.6401$ & $-1.4129$ & $-1.7130$ & $-1.5363$ & $-1.7866$ & $-1.6240$ & $-1.6829$ & $-1.5351$ & $-1.7126$ & $-1.5696$ \\
  $^{86}$Zn   & $\beta^-$ & $-2.2042$ &           & $-2.3010$ &           & $-2.2781$ & $-2.1195$ & $-2.3513$ & $-2.2053$ & $-2.3934$ & $-2.3172$ & $-2.4593$ & $-2.3885$ & $-2.3631$ & $-2.2542$ & $-2.4675$ & $-2.3709$ \\
  $^{92}$Ge   & $\beta^-$ & $-1.9708$ & $-2.0242$ & $-1.9664$ & $-2.0202$ & $-2.0997$ & $-2.0801$ & $-2.1400$ & $-2.1213$ & $-2.2194$ & $-2.2592$ & $-2.1557$ & $-2.1983$ & $-2.1844$ & $-2.1605$ & $-2.2128$ & $-2.1896$ \\
  $^{142}$Er  & $\beta^+$/EC & $-1.5552$ & $-1.5435$ & $-1.5681$ & $-1.5579$ & $-1.5045$ & $-1.4746$ & $-1.5327$ & $-1.5105$ & $-1.5674$ & $-1.5632$ & $-1.5738$ & $-1.5699$ & $-1.5697$ & $-1.5690$ & $-1.5741$ & $-1.5735$ \\
  $^{194}$W   & $\beta^-$ & $1.7800$  &           & $1.7726$  &           & $0.8705$  & $2.0612$  & $0.5726$  & $1.4293$  & $0.4638$  & $1.0065$  & $0.2473$  & $0.6840$  & $0.5495$  & $0.9044$  & $0.5350$  & $0.8852$  \\
  \bottomrule
  \end{tabular}%
  }
  \end{table*}
Since our predictions demonstrate good agreement with experimental half-lives and branching ratios, we extend the calculations to nuclei where experimental information is absent or highly uncertain: $^{86}$Zn, $^{92,94}$Se, $^{92}$Ge, $^{142}$Er, $^{192}$Hf, and $^{194}$W. The selenium isotopes reside in the $A \approx 90$ region traversed by rp-process flows in X-ray bursts~\cite{schatz1998rp}; $^{142}$Er is relevant to p and $\nu$p-process nucleosynthesis~\cite{arcones2011review}; and $^{192}$Hf and $^{194}$W lie near the $N=126$ shell closure where $\beta$-decay lifetimes directly control the third $r$-process abundance peak~\cite{mumpower2016impact, surman2014sensitivity}.

The predicted half-lives are summarized in Table~\ref{tab:logt_half}, where results from all four deformation-sensitive models (NSM, NL3$^*$, DD-ME2, DD-PC1) are presented across all transition pathways. A recurring pattern is that variations in $Q$-values along different pathways producing measurable changes in predicted half-lives, with $\log_{10}T_{1/2}$ values spanning up to $\sim$0.3s-1.0s, for nuclei exhibiting pronounced shape coexistence. This sensitivity arises from $Q^{-1/4}$ dependence of the phase-space factor: even fractionally small shifts in $Q$ between pathways propagate through this inverse-power dependence to yield measurable changes in $\log_{10}T_{1/2}$. For $^{92}$Se, our models span $\log_{10}T_{1/2} \approx -0.93$s to $-1.29$s across all pathways. In $^{194}$W, the pathway dependence is particularly pronounced: NSM yields a narrow band of 1.77s--1.78s, while NL3$^*$, DD-ME2, and DD-PC1 span a much wider range of $\approx$0.25s-2.06s, underscoring how strongly the predicted half-life depends on the structural configuration of both parent and daughter. For $^{86}$Zn, all four models consistently predict half-lives in the range $\approx -2.20$s to $-2.47$s, with the spread directly quantifying the structural changes introduced by shape coexistence. Our analysis reveals two distinct sources of uncertainty in half-life predictions for shape-coexisting nuclei. The dominant contribution arises from pathway dependence, the choice of which energy minimum the parent and daughter nuclei occupy, spanning 0.3s to 1.0s in $\log_{10}T_{1/2}$. This pathway-dependent uncertainty directly reflects the physical reality of shape coexistence and becomes particularly significant at key $r$-process waiting-point nuclei, where precise half-life predictions are crucial for nucleosynthesis modeling. A secondary, smaller contribution comes from parametric uncertainty inherent in the deformation-dependent decay formulas themselves, amounting to $\sim$0.04s-0.05s in $\log_{10}T_{1/2}$. A fully microscopic treatment of configuration mixing and electromagnetic decay widths between coexisting minima would require the explicit calculation of transition matrix elements within a beyond-mean-field framework, such as a self-consistent QRPA or GCM-based approach. The present study is intentionally restricted to a deformation-dependent phenomenological analysis of decay systematics. An extension towards a unified description incorporating configuration mixing, electromagnetic transition strengths and microscopic Gamow-Teller distributions represents a natural direction for future investigation.

\section{Conclusion}
This work demonstrates the significant influence of shape coexistence on nuclear decay properties. Shape coexistence, where competing intrinsic configurations exist at nearly degenerate energies, directly affects both half-lives and branching ratios by altering the structural overlap between parent and daughter states. The study presents a survey across the nuclear chart, combining the Nilsson–Strutinsky Method (NSM) with relativistic mean-field (RMF) calculations employing the NL3$^*$, DD-ME2, and DD-PC1 covariant energy density functionals. From approximately 400 even-even nuclei predicted to exhibit shape coexistence, we demonstrate how transitions between competing configurations in parent and daughter nuclei impact decay behavior. The results establish that shape coexistence plays a crucial role in decay systematics. When both parent and daughter nuclei can occupy multiple energy minima, the choice of transition pathway substantially alters predicted lifetimes, with half-lives varying by factors of 1.35s to 2.46s depending on the nucleus. Shape coexistence also influences branching ratios, with certain nuclei exhibiting complete inversions of the dominant decay mode depending on which configurations participate in the transition. These configuration-dependent variations demonstrate that shape coexistence is not merely a structural curiosity but a physical reality that measurably affects nuclear stability and decay dynamics. Our findings establish that deformation and its associated uncertainties are essential for reliable decay predictions in transitional regions. Even modest structural changes can significantly alter decay rates and branching ratios, with important implications for astrophysical nucleosynthesis where half-life uncertainties directly impact abundance patterns. The framework developed here, incorporating deformation effects into decay calculations, provides an approach for predicting lifetimes in regions where competing nuclear shapes coexist. As experimental facilities continue to extend measurements toward exotic nuclei, the shape-coexisting candidates identified in this study serve as valuable benchmarks for testing theoretical predictions and refining nuclear models relevant to both nuclear structure and astrophysics.

\vspace*{5pt}
\section*{Acknowledgments}
The authors are grateful to Prof. Nils Paar, University of Zagreb, Croatia for his valuable help with the calculations of DDME2 and DDPC1. G.S. and N.C. express gratitude for the support received from the Department of Science \& Technology (DST), Government of Rajasthan, Jaipur, India under F24(1)DST/R\&D/2024-EAC-00378-6549873/819. SP acknowledges the funding from Severo Ochoa Center of Excellence under the CEX2023-001292-S grant, funded by MCIN/AEI/10.13039/501100011033. SP extends his gratitude to the Institute Jožef Stefan (IJS), Ljubljana, for providing access to computational resources that supported the theoretical part of this work. Special thanks are extended to Prof. Rok Pestotnik (IJS, Ljubljana) and Prof. Peter Križan (IJS, Ljubljana) for their support. M.A. and P.P. thanks DST-Govt. of India for the support under WIDUSHI-B/PM/2024/23.

\bibliographystyle{apsrev4-2}

\end{document}


\title{Supplementary Information\\ Impact of Shape Coexistence on Nuclear Stability}

\author{G.~Saxena}
\author{H.~Sikhwal}
\author{N.~Chandnani}
\author{Pranali~Parab}
\author{Siddharth~Parashari}
\author{Gabriela~Llosá}
\author{Mamta~Aggarwal}

\maketitle

\section{Theoritical Formalism}
\subsection{Nilsson Strunsiky Method}
The shape of an atomic nucleus is influenced by the subtle balance between macroscopic bulk effects of nuclear matter and microscopic shell effects for which we employ the triaxially deformed Nilsson model, along with the Strutinsky prescription, beginning with the Strutinsky density distribution function ~\cite{Strutinsky1967, BrackDamgaard1972,nilsson1969new} for single particle states as
\begin{equation}
\tilde {g(\epsilon)} = \frac{1}{\sqrt{\pi}\gamma} \sum exp(-u_i)^2 \sum_{k=0}^{\infty} C_k H_k (u_i),
\end{equation}
where
\begin{equation}
u_i = \frac{\epsilon - \epsilon_i}{\gamma},
\end{equation}
and the coefficients C$_k$ are
\begin{equation}
    C_{k} = \begin {cases}
    \frac{(-1)^{\frac{k}{2}}}{2^{k}\frac{k}{2}!}, \hspace{0.2cm} \forall \hspace{0.1cm} even \hspace{0.1cm} k;\\
   0, \hspace{0.9cm} \forall \hspace{0.1cm} odd \hspace{0.1cm} k.
    \end{cases}
\end{equation}
Hermite polynomials H$_k$(u$_i$) up to high order of correction are utilized to create a smooth distribution of levels. The energy obtained from Strutinsky's smoothened single particle level distribution is expressed as
\begin{equation}
\tilde E = \int_{-\infty}^{\mu} \tilde g(\epsilon) d\epsilon
\end{equation}
The chemical potential $\mu$ is fixed by the number conserving equation
\begin{equation}
N = \int_{-\infty}^{\mu} \tilde g(\epsilon) d\epsilon
\end{equation}
The shell correction to the energy is obtained as
\begin{equation}
\delta E = \sum_{i=1} ^A \epsilon_i- \tilde E
\end{equation}
The first term shows the energy of the shell model in the ground state, while the latter term is the smoothed energy containing the smearing width 1.2$\hbar$$\omega$.
The Nilsson Hamiltonian generates the single particle energies $\epsilon_i$ as a function of deformation parameters ($\beta$, $\gamma$) for the deformed oscillator diagonalized in the cylindrical representation ~\cite{shanmugam1979nuclear,eisenberg1972w}.
\begin{eqnarray}
H=p^2 /2m + (m/2)(\omega_x^2 x^2 + \omega_y^2 y^2 +\omega_z^2 z^2)+  \nonumber\\
 C l.s + D (l^2- 2<l^2>)
\end{eqnarray}
The coefficients for the $l.s $ and $l{^2}$ terms are taken from Seeger~\cite{eisenberg1972w} who has fitted them to reproduce the shell corrections~\cite{Strutinsky1967} to ground state masses. Strutinsky's shell correction to energy $\delta$E added to macroscopic energy of the spherical drop $BE_{LDM}$ reproduced by LDM mass formula~\cite{moller1995data} along with the deformation energy $E_{def}$ (obtained from surface and coulomb effects) gives the total energy $BE_{gs}$ corrected for microscopic effects of the nuclear system
\begin{eqnarray}
BE_{gs}(Z,N,\beta,\gamma) = BE_{LDM}(Z,N) -  \nonumber\\
 E_{def}(Z,N,\beta,\gamma)-\delta E_{shell}(Z,N,\beta,\gamma)
\end{eqnarray}

The Nilsson-Strutinsky calculations were performed using an in-house implementation of the method following the formalism described above, 
with Nilsson parameters taken from Ref. \cite{SEEGER1975491} and liquid-drop mass formula parameters from Ref. \cite{moller1995data}. In order to examine the equilibirum deformation and nuclear shapes, energy $E (= -BE)$ minima are searched for a variety of $\beta$ and $\gamma$ with axial deformation parameter and $\gamma$ ranging from $-180^o$ (oblate) to $-120^o$ (prolate) and triaxial in between. The exact position of the first unbound proton and neutron is located by one of the proton/neutron separation energies tending to zero value, which we determine by calculating the difference in binding energy $BE_{gs}$ based on the binding of both the parent and daughter nucleus.

\subsection{Relativistic Mean-Field (RMF) model}
The present study focuses on the approach of Relativistic Mean field (RMF) where the different variants of the RMF models have been used for the calculations \cite{lalazissis2009effective}. The first variant of the RMF comprises of Langrangian density that include non linear terms for the $\sigma$ and $\omega$ mesons: 
\begin{eqnarray}
       {\cal L}& = &{\bar\psi} [\imath \gamma^{\mu}\partial_{\mu}
                  - M]\psi\nonumber\\
                  &&+ \frac{1}{2}\, \partial_{\mu}\sigma\partial^{\mu}\sigma
                - \frac{1}{2}m_{\sigma}^{2}\sigma^2- \frac{1}{3}g_{2}\sigma
                 ^{3} - \frac{1}{4}g_{3}\sigma^{4} -g_{\sigma}
                {\bar\psi}  \sigma  \psi\nonumber\\
               &&-\frac{1}{4}H_{\mu \nu}H^{\mu \nu} + \frac{1}{2}m_{\omega}
                  ^{2}\omega_{\mu}\omega^{\mu} + \frac{1}{4} c_{3}
                 (\omega_{\mu} \omega^{\mu})^{2}
                  - g_{\omega}{\bar\psi} \gamma^{\mu}\psi
                 \omega_{\mu}\nonumber\\
              &&-\frac{1}{4}G_{\mu \nu}^{a}G^{a\mu \nu}
                 + \frac{1}{2}m_{\rho}
                 ^{2}\rho_{\mu}^{a}\rho^{a\mu}
                  - g_{\rho}{\bar\psi} \gamma_{\mu}\tau^{a}\psi
                 \rho^{\mu a}\nonumber\\
               &&-\frac{1}{4}F_{\mu \nu}F^{\mu \nu}
                 - e{\bar\psi} \gamma_{\mu} \frac{(1-\tau_{3})}
                 {2} A^{\mu} \psi\,\,,
\end{eqnarray}

where the field tensors $H$, $G$ and $F$ for the vector fields are defined by
\begin{eqnarray}
                 H_{\mu \nu} &=& \partial_{\mu} \omega_{\nu} -
                       \partial_{\nu} \omega_{\mu}\nonumber\\
                 G_{\mu \nu}^{a} &=& \partial_{\mu} \rho_{\nu}^{a} -
                       \partial_{\nu} \rho_{\mu}^{a}
                     -2 g_{\rho}\,\epsilon^{abc} \rho_{\mu}^{b}
                    \rho_{\nu}^{c} \nonumber\\
                  F_{\mu \nu} &=& \partial_{\mu} A_{\nu} -
                       \partial_{\nu} A_{\mu}\,\,\nonumber\
\end{eqnarray}

Other symbols have their usual meaning and further details can be found in Refs.~\cite{boguta1977relativistic,boguta1983systematics, furnstahl1997chiral, geng2003relativistic}
For the calculations in this work, we have started with first variant of the RMF model whch is NL3$^*$ \cite{lalazissis2009effective}. It includes linear terms for the $\sigma$, $\omega$ and $\rho$ mesons and inclusion of non-linear term for the self-interaction of the $\sigma$ meson only. In these non-linear versions of our calculations, we use a delta force, i.e., $V = -V_{0} \delta(r)$ with the strength $V_{0} = 350 MeV fm^3$ for the pairing interaction, which has been used in Refs. \cite{yadav2004description, saxena2019bubble} for the successful description of drip-line nuclei. Apart from its simplicity, the applicability and justification of using such a $\delta$-function form of interaction has been discussed in Ref. \cite{dobaczewski1984hartree}, whereby it has been shown in the context of HFB calculations that the use of a delta force in a finite space simulates the effect of finite range interaction in a phenomenological manner (see Ref.~\cite{bertsch1991pair} for more details). The computer code used for these calculations is given by Ring \textit{et al.} \cite{ring1997program}. The parameter set used is NL3$^*$ \cite{lalazissis2009effective}, which has been widely used to provide an excellent description of spherical as well as in deformed nuclei. For further details of these formulations we refer the reader to Refs. \cite{boguta1977relativistic,boguta1983systematics,furnstahl1997chiral,lalazissis2009effective,singh2012study,geng2003relativistic}.

The second and third variant of the RMF model belongs to the category of density dependence of coupling constants for the meson exchange that includes DD-ME2 \cite{lalazissis2005new} and DD-PC1 \cite{NIKSIC20141808}. Calculations for both functionals were performed using the DIRHB (Dirac-Hartree-Bogoliubov) code developed by Nik\v{s}i\'{c}, Vretenar, and Ring \cite{nikvsic2014dirhb}.

For the density dependent model of meson exchange (DD-ME), interaction part does not contain any non-linear term, but includes the strengths of meson-nucleon couplings $g_\sigma$, $g_\omega$, and $g_\rho$ which have a density dependence explicitly in the following form:
\begin{equation}
g_i(\rho) = g_i(\rho_{\text{sat}})\, f_i(x), \qquad \text{for } i = \sigma, \omega
\end{equation}

where the density dependence is given by

\begin{equation}
f_i(x) = a_i \frac{1 + b_i (x + d_i)^2}{1 + c_i (x + d_i)^2}
\end{equation}

in which $x$ is given by $x = \rho / \rho_{\text{sat}}$, and $\rho_{\text{sat}}$ denotes the baryon density at saturation in symmetric nuclear matter. For the $\rho$ meson, density dependence is of exponential form and given by

\begin{equation}
f_\rho(x) = \exp\!\left(-a_\rho (x - 1)\right)
\end{equation}

The effective Langragain defined for DD-PC1 model is analogous to DD-ME2 model but, it does not include derivative term for mesonic field and hence they are directly expressed in terms of nucleonic field
\begin{eqnarray}
\mathcal{L}_{\text{int}} &=&
-\frac{1}{2}\alpha_S(\rho)(\bar{\psi}\psi)(\bar{\psi}\psi)
-\frac{1}{2}\alpha_V(\rho)(\bar{\psi}\gamma^\mu\psi)(\bar{\psi}\gamma_\mu\psi) \nonumber\\
&&
-\frac{1}{2}\alpha_{TV}(\rho)(\bar{\psi}\vec{\tau}\gamma^\mu\psi)(\bar{\psi}\vec{\tau}\gamma_\mu\psi) \nonumber\\
&&
-\frac{1}{2}\delta_S(\rho)(\bar{\psi}\psi)\Box(\bar{\psi}\psi)
\end{eqnarray}

In analogy with the meson-exchange model (DD-ME) which has shown above, this model
contains interactions of isoscalar-scalar (S), isoscalar-vector (V) and isovector-vector (TV). The coupling constants $\alpha_i(\rho)$ are density dependent and have the form \cite{35}:

\begin{equation}
\alpha_i(\rho) = a_i + (b_i + c_i x)e^{-d_i x}, 
\qquad \text{for } i = S, V, TV
\end{equation}

For both frameworks, the DIRHB code employs a separable pairing force in the $T = 1$ channel based on a finite-range interaction of Gogny type \cite{tian2009finite}. In coordinate space, the pairing force reads:
\begin{equation}
V(\mathbf{r}_1, \mathbf{r}_2, \mathbf{r}_1', \mathbf{r}_2') = G \, \delta(\mathbf{R} - \mathbf{R}') \, P(\mathbf{r}) \, P(\mathbf{r}')
\end{equation}
where $\mathbf{R} = (\mathbf{r}_1 + \mathbf{r}_2)/2$ is the center-of-mass coordinate, $\mathbf{r} = \mathbf{r}_1 - \mathbf{r}_2$ is the relative coordinate, and
\begin{equation}
P(\mathbf{r}) = \frac{1}{(4\pi a^2)^{3/2}} \exp\left(-\frac{r^2}{4a^2}\right)
\end{equation}
is a Gaussian form factor with strength parameter $G = 728 \text{ MeV fm}^3$ and range $a = 0.644 \text{ fm}$. This formulation, adjusted to reproduce the pairing gap of the Gogny D1S force in symmetric nuclear matter \cite{tian2009finite}.

These relativistic EDF parameterizations like DD-ME2 and DD-PC1 are mainly optimized for even-even nuclei; their deformation predictions do not explicitly cover odd-A or odd-odd systems, making their direct use for odd nuclei formally inconsistent. However this limitation is not unique, Skyrme-EDF global $\beta$-decay studies include odd nuclei via approximations like the equal-filling approximation, which preserves time-reversal symmetry by equally distributing the unpaired nucleon over a Kramers pair~\cite{shafer2016beta,ney2020global}. In the present work, Q-values are computed from energy differences between parent and daughter nuclei, reducing irregularities in odd-nucleus treatment and enabling meaningful exploration of shape coexistence within a deformation-sensitive framework.

\section{Numerical Details and Convergence}

\begin{figure*}[htbp]
    \centering
    \includegraphics[width=\textwidth]{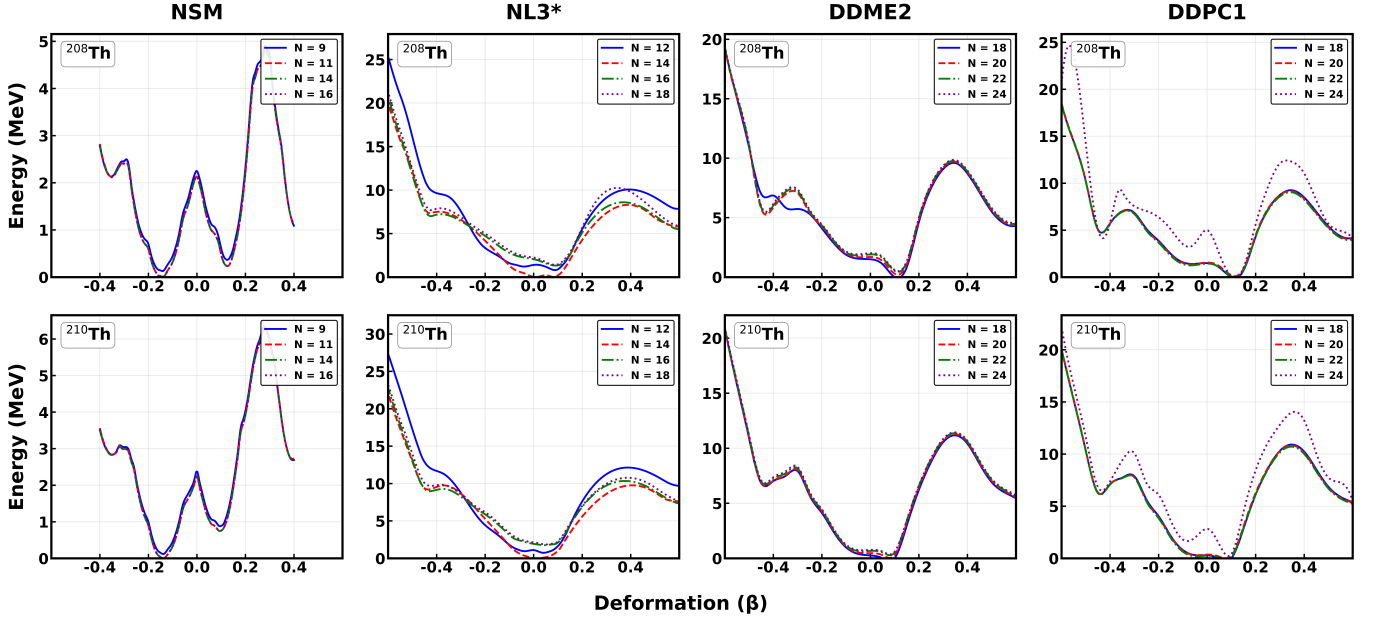}
    \caption{Convergence of the potential energy surface for the isotopes $^{208}$Th and $^{210}$Th as a function of the quadrupole deformation parameter $\beta$ for the NSM, NL3$^{*}$, DD-ME2, and DD-PC1 models. The different curves in each panel correspond to increasing basis sizes (number of oscillator shells, $N$). The energy landscapes, and importantly, the relative energy differences between the first and second minima, demonstrate clear convergence at higher $N$ values.}
    \label{fig:convergence}
\end{figure*}

Two important parameters in the harmonic oscillator basis expansion are 
$N_f$ and $N_b$, the number of shells used to expand the fermion and 
boson fields, respectively. The models employed in this work, NL3$^{*}$, 
DD-ME2, DD-PC1, and the NSM have been extensively 
validated across the nuclear chart, from light nuclei to the superheavy 
region. In particular, Geng, Toki, and Meng~\cite{geng2005masses} 
performed a systematic study of ground-state properties for 6969 nuclei 
spanning $Z \leq 100$ from the proton to the neutron drip line using 
the RMF model, demonstrating that $N_f = N_b = 12$ is sufficient for 
light and medium-mass nuclei while larger spaces are required for heavier 
systems. For the density-dependent functionals DD-ME2 and DD-PC1, 
Agbemava et al.~\cite{agbemava2014global} performed large-scale 
relativistic Hartree-Bogoliubov calculations for all $Z \leq 104$ 
even-even nuclei using $N_f = N_b = 20$, establishing this as the 
standard basis for global calculations with these functionals. The same 
three functionals, DD-ME2, DD-PC1, and NL3$^{*}$, were further applied with 
the same basis sizes to a comprehensive study of ground-state and fission 
properties of actinides and superheavy nuclei with $Z = 90$-$120$ 
throughout the drip lines~\cite{agbemava2020cdft}.

To verify the stability of the potential energy surface topology 
and the deformation parameters across different basis sizes, convergence 
tests have been carried out for the representative actinide nuclei 
$^{208}$Th and $^{210}$Th using all four theoretical models. These nuclei 
are selected as benchmark cases since they lie in the heavy-mass region 
directly relevant to the present study and exhibit shape coexistence 
between competing prolate and oblate configurations, making them 
particularly sensitive to basis truncation effects. The basis sizes 
adopted in this work, $N = 12$ and $N = 11$ for NL3$^{*}$ and NSM, respectively, as well as, $N = 20$ for 
DD-ME2 and DD-PC1 are consistent with well-established global 
calculations spanning light to superheavy nuclei 
\cite{geng2005masses, agbemava2014global, agbemava2020cdft}.

For the three relativistic mean-field models (NL3$^{*}$, DD-ME2, DD-PC1), calculations are performed in a harmonic oscillator basis where nucleon spinors and meson fields are expanded in axially symmetric deformed harmonic oscillator states. The convergence is studied by systematically increasing the number of major oscillator shells $N = N_f = N_b$. For NL3$^{*}$, we tested $N = 12, 14, 16, 18$, where $N = 12$ is our production setting. For DD-ME2 and DD-PC1, which employ the DIRHB solver with separable pairing, we tested $N = 18, 20, 22$ and $24$ where $N = 20$ is our production basis. For the macroscopic-microscopic Nilsson-Strutinsky method (NSM), single-particle states are calculated by diagonalizing the Nilsson Hamiltonian in a cylindrical harmonic oscillator basis. We tested $N = 9, 11, 14, 16$ shells, with $N = 11$ used for the main calculations.

Figure~\ref{fig:convergence} displays the normalized binding energy curves as a function of quadrupole deformation $\beta$ for both test nuclei across all four models. For NSM, the curve convergences at all considered values ($N=9, 11, 14, 16$). In fact, the convergence is remarkably rapid: already at $N = 11$ the PES is indistinguishable from larger basis calculations, and $N = 14, 16$ produce numerically identical results. For NL3$^{*}$, the curves systematically converge as the basis increases, with stabilization achieved by $N = 16$ and $N = 18$. For DD-ME2 and DD-PC1, the energy landscapes exhibit stability, with curves for $N = 18, 20, 22$ and $24$ overlapping almost perfectly, demonstrating that $N = 20$ is more than sufficient for these density-dependent functionals.

\begin{table}[htbp]
\centering
\tiny
\caption{Deformation parameters $\beta_1$ (prolate minimum) and 
$\beta_2$ (oblate minimum) for $^{208}$Th and $^{210}$Th at different 
basis sizes $N$ for all four models. $\Delta B$ (MeV) is the binding 
energy deviation relative to the production basis (indicated in bold).}
\label{tab:minima}
\setlength{\tabcolsep}{4pt}
\resizebox{\columnwidth}{!}{%
\begin{tabular}{llcccc}
\toprule
Nucleus & Model & $N$ & $\beta_1$ & $\beta_2$ & $\Delta B$ (MeV) \\
\midrule

\multirow{16}{*}{$^{208}$Th}
& NSM    & 9  & $-0.140$ & $+0.120$ & $0.118$ \\
&        & \textbf{11} & $\mathbf{-0.140}$ & $\mathbf{+0.130}$ & \textbf{0.000} \\
&        & 14 & $-0.140$ & $+0.130$ & $0.009$ \\
&        & 16 & $-0.140$ & $+0.130$ & $0.009$ \\
\cmidrule{2-6}

& NL3$^{*}$ & \textbf{12} & $\mathbf{+0.086}$ & $\mathbf{-0.042}$ & \textbf{0.000} \\
&           & 14 & $+0.083$ & $-0.035$ & $0.783$ \\
&           & 16 & $+0.096$ & $-0.043$ & $0.527$ \\
&           & 18 & $+0.094$ & $-0.039$ & $0.635$ \\
\cmidrule{2-6}

& DDME2 & 18 & $+0.100$ & $-0.050$ & $0.284$ \\
&       & \textbf{20} & $\mathbf{+0.100}$ & $\mathbf{-0.050}$ & \textbf{0.000} \\
&       & 22 & $+0.100$ & $-0.050$ & $0.239$ \\
&       & 24 & $+0.100$ & $-0.050$ & $0.321$ \\
\cmidrule{2-6}

& DDPC1 & 18 & $+0.100$ & $-0.050$ & $0.190$ \\
&       & \textbf{20} & $\mathbf{+0.100}$ & $\mathbf{-0.050}$ & \textbf{0.000} \\
&       & 22 & $+0.100$ & $-0.050$ & $0.108$ \\
&       & 24 & $+0.100$ & $-0.100$ & $0.070$ \\
\midrule

\multirow{16}{*}{$^{210}$Th}
& NSM    & 9  & $-0.140$ & $+0.100$ & $0.117$ \\
&         & \textbf{11} & $\mathbf{-0.140}$ & $\mathbf{+0.100}$ & \textbf{0.000} \\
&        & 14 & $-0.140$ & $+0.100$ & $0.002$ \\
&        & 16 & $-0.140$ & $+0.100$ & $0.002$ \\
\cmidrule{2-6}

& NL3$^{*}$ & \textbf{12} & $\mathbf{+0.037}$ & $\mathbf{-0.035}$ & \textbf{0.000} \\
&           & 14 & $+0.039$ & $-0.030$ & $0.655$ \\
&           & 16 & $+0.056$ & $-0.037$ & $1.034$ \\
&           & 18 & $+0.055$ & $-0.035$ & $1.113$ \\
\cmidrule{2-6}

& DDME2 & 18 & $+0.100$ & $-0.050$ & $0.288$ \\
&       & \textbf{20} & $\mathbf{+0.050}$ & $\mathbf{-0.050}$ & \textbf{0.000} \\
&       & 22 & $+0.100$ & $-0.050$ & $0.234$ \\
&       & 24 & $+0.100$ & $-0.050$ & $0.307$ \\
\cmidrule{2-6}

& DDPC1 & 18 & $+0.050$ & $-0.050$ & $0.163$ \\
&       & \textbf{20} & $\mathbf{+0.050}$ & $\mathbf{-0.050}$ & \textbf{0.000} \\
&       & 22 & $+0.050$ & $-0.050$ & $0.108$ \\
&       & 24 & $+0.100$ & $-0.100$ & $0.232$ \\
\bottomrule
\end{tabular}}
\end{table}

Table~\ref{tab:minima} quantifies this stability by listing the 
deformation parameters $\beta_1$ and $\beta_2$ of the prolate and 
oblate minima, along with the binding energy deviation $\Delta B$ 
relative to the production basis for each model. The most important 
observation is that the deformation parameters $\beta_1$ and $\beta_2$ 
remain nearly unchanged across all values of $N$ for all four 
models. For NSM, the prolate and oblate deformations are completely 
stable from $N = 9$ onward. For NL3*, both $\beta_1$ and $\beta_2$ 
vary by at most 0.02 units across $N = 12$--$18$. For DD-ME2 and 
DD-PC1, the deformation parameters are invariant across $N = 18$--$24$. 
This consistency in the predicted shapes across different basis sizes 
demonstrates that the shape coexistence features identified in this 
work are physically meaningful and not artifacts of basis truncation.

\section{Half-Life Calculations}
Theoretical treatments for estimating $\beta$ and $\alpha$-decay lifetimes span a range of models, each tailored to capture different aspects of nuclear dynamics. Several semi-empirical models exist for estimating half-life across all decay modes. The predictive performance of any empirical formula depends on the dataset over which it is evaluated; consequently, RMSE values quoted in the original publications cannot be directly compared across studies because they obtained from different nuclear datasets, fitting procedures, and selection criteria. To ensure a consistent and fair comparison, all benchmark RMSE and reduced $\chi^2$ values reported here have been recomputed on the same datasets used in the present work: 423 $\alpha$-decay nuclei, 1046 $\beta^{-}$ emitters, and 955 $\beta^{+}/EC$  nuclei from NUBASE2020.

\subsection{Performance Metrics}
The predictive accuracy of all formulas is quantified using two statistical approaches: the root-mean-square error (RMSE) and the reduced chi-square statistic ($\chi^2$):
\begin{align}
\mathrm{RMSE} &= \sqrt{\frac{1}{N_d} \sum_{i=1}^{N_d}
\left(\log T_{\mathrm{Th}}^i - \log T_{\mathrm{Expt.}}^i\right)^2},
\label{eq:rmse} \\
\chi^2 &= \frac{1}{N_d - N_p} \sum_{i=1}^{N_d}
\left(\log T_{\mathrm{Th}}^i - \log T_{\mathrm{Expt.}}^i\right)^2,
\label{eq:chi2}
\end{align}
where $N_d$ is the number of data points and $N_p$ is the number of fitted parameters.

These quantities measure, the average logarithmic deviation between theoretical and experimental half-lives and the goodness-of-fit accounting for the number of fitted parameters. 

\subsection{Proposed Formulas}

Existing $\beta$-decay and $\alpha$-decay half-lives models perform well in reproducing global decay trends but, they do not explicitly account for nuclear deformation or structural symmetry between parent and daughter nuclei. This limitation becomes especially significant in regions where shape coexistence occurs, as transitions between prolate and oblate configurations often separated by small energy differences can substantially affect decay probabilities. To address this, we propose a deformation-sensitive extension to investigate whether such structural effects contribute meaningfully to the behavior of half-lives. Following the methodology outlined in Jain \textit{et al.} \cite{jain2024shape_scientific_reports}, we introduce an additional term dependent on the quadrupole deformation parameters of the parent and daughter nuclei, denoted by $\beta_p$ and $\beta_d$, respectively. This modification has previously demonstrated success in estimating the half-lives of proton-emitting nuclei \cite{jain2024shape_scientific_reports}, and in the present context, it plays a crucial role by incorporating the influence of nuclear shape transitions. Given that small energy fluctuations can lead to changes in nuclear shape, this deformation-dependent correction becomes both necessary and physically well-motivated.

For $\alpha$-decay, the proposed formula is:
\begin{equation}
\log_{10}\left(T_{1/2}\right)
=
c_0\,\frac{Z_d}{\sqrt{Q}} + c_1\,\sqrt{Z\,A^{1/3}} + c_2\,\mathrm{D} + c_3
\label{eq:alpha_new}
\end{equation}
where $Z_d$ is the atomic number of the daughter nucleus, $Q$ is the $\alpha$-decay energy in MeV, $Z$ and $A$ are the atomic and mass numbers of the parent nucleus, and $D$ is the deformation term.

For $\beta^{-}$ and $\beta^{+}$ decay, the proposed formula is:
\begin{equation}
\log_{10}\left(T_{1/2}\right)
=
c_0 + c_1\,\delta + c_2\,\mathrm{I}
+ c_3\,Q^{-1/4}
+ c_4\,\frac{Z}{A^{1/3}}
+ c_5\,\mathrm{D}
\label{eq:betaminus_new}
\end{equation}
where $\delta$ is the pairing term, $\mathrm{I} = (N-Z)/A$ is the isospin asymmetry, and $Z/A^{1/3}$ serves as a Coulomb proxy and D is the deformation term. The $Q^{-1/4}$ term captures the inverse correlation between half-life and available decay energy.

In both Eq~(\ref{eq:alpha_new}) and Eq~(\ref{eq:betaminus_new}), the half-life is measured in seconds, and the deformation term is:
\begin{equation}
D = (-1)^{\kappa_p + \kappa_p\kappa_d} \left( \kappa_p \beta_p \right)^{1/2}+(-1)^{\kappa_d + \kappa_p\kappa_d} \left( \kappa_d \beta_d \right)^{1/2}
\label{eq:deformation}
\end{equation}

Here, $\beta_p$ and $\beta_d$ are the quadrupole deformation parameters of the parent and daughter nuclei, and $\kappa$ denotes a shape parameter assigned as $+2$ for prolate and $-1$ for oblate shapes, following the approach proposed in Ref~\cite{jain2024shape_scientific_reports}.
The exponential factor $(-1)^{\kappa_p + \kappa_d}$ is designed to enhance transitions between shape-similar nuclei (e.g., prolate $\rightarrow$ prolate) and penalize shape-changing transitions, reflecting their higher structural rearrangement costs \cite{taageper1961relations}. 

In absence of experimental quadrupole deformation parameters of several considered nuclei of the datasets, to accurately assign the $\beta_p$ and $\beta_d$ in the modified formula, we require consistent and reliable theoretical estimates of nuclear deformation for both parent and daughter nuclei. For this purpose, we extract values from four independent nuclear structure models: the Finite Range Droplet Model (FRDM) \cite{moller2016nuclear}, the Hartree-Fock-Bogoliubov model (HFB) \cite{ebran_khan_peña_arteaga_vretenar_2011}, the Weizs\"{a}cker-Skyrme-4 (WS4) \cite{Zhao_2019} and the Relativistic Mean Field (RMF) approach \cite{10.1143/PTP.110.921}. Each of these models predict $\beta$ based on distinct assumptions and energy functionals. However, relying on a single theoretical model can introduce systematic bias, especially in regions where theoretical predictions diverge from one another or where data (especially deformation data) is limited.
To address this, we employed a more robust approach, Bayesian Model Averaging (BMA), to combine the deformation values from the four theoretical models. BMA is a statistical framework that weights each model using posterior probability based on its ability to reproduce known experimental observables \cite{höge_guthke_nowak_2020}. The model-averaged deformation parameter represents an agreed-upon estimate that takes into account both variation between models and available empirical constraints.

\subsection{Fitted Coeffcients}
The final fitting performed via linear least squares estimation. Paramaeter uncertanities are propogated to provide confidence intervals for each model coefficient, serving both as a measure of statistical reliability and as a basis for predictive intervals when the formulas are applied to unknown or exotic nuclei.

For the $\alpha$-decay mode, the fitted coefficients are
$c_0 = 1.6148 \pm 0.0017$,
$c_1 = -1.1584 \pm 0.0038$,
$c_2 = -0.0193 \pm 0.0140$, and
$c_3 = -24.7221 \pm 0.1041$.

For $\beta^{-}$ decay, the extracted parameters are
$c_0 = -2.5172 \pm 0.0024$,
$c_1 = -0.4124 \pm 0.0035$,
$c_2 = -14.2754 \pm 0.0678$,
$c_3 = 9.1383 \pm 0.0184$,
$c_4 = 0.0665 \pm 0.0012$, and
$c_5 = -0.1948 \pm 0.0040$.

For $\beta^{+}$/EC decay, the fitted values are
$c_0 = 8.2980 \pm 0.0010$,
$c_1 = 0.3512 \pm 0.0019$,
$c_2 = 47.3990 \pm 0.0483$,
$c_3 = 0.8393 \pm 0.0064$,
$c_4 = -0.9717 \pm 0.0012$, and
$c_5 = -0.0196 \pm 0.0021$.

Table~\ref{tab:rmse_chi2} summarises the RMSE and reduced $\chi^2$ for the present
formula and all benchmark models evaluated on the complete datasets
(423 $\alpha$-decay, 1046 $\beta^{-}$, and 955 $\beta^{+}$/EC nuclei).
All RMSE and $\chi^2$ values are computed on the
same datasets to ensure a consistent comparison. 

\begin{table}[h]
\centering
\caption{RMSE and reduced $\chi^2$ for the present formula and benchmark models on the complete datasets, arranged by RMSE (lowest to highest). Present formula results are highlighted in bold.}
\label{tab:rmse_chi2}
\small
\begin{tabular}{llcc}
\toprule
Decay Mode & Formula & RMSE & $\chi^2$ \\
\midrule
\multirow{5}{*}{$\alpha$}
& \textbf{Present work} & \textbf{0.635} & \textbf{0.407} \\
& IUF2022~\cite{ismail2022improved} & 0.658 & 0.440 \\
& Royer2020~\cite{royer2020} & 0.693 & 0.490 \\
& QF2021~\cite{saxena2021new} & 0.904 & 0.826 \\
& NMHF2021~\cite{sharma2021new} & 0.955 & 0.927 \\
\midrule
\multirow{4}{*}{$\beta^{-}$}
& \textbf{Present work} & \textbf{0.717} & \textbf{0.517} \\
& Sobhani~(2022)~\cite{sobhani2023beta} & 0.773 & 0.601 \\
& Zhou~(2017)~\cite{zhou2017empirical} & 0.834 & 0.702 \\
& Zhang~(2007)~\cite{zhang2007systematics}& 1.158 & 1.349 \\
\midrule
\multirow{2}{*}{$\beta^{+}$/EC}
& \textbf{Present work} & \textbf{0.968} & \textbf{0.943} \\
& Sobhani~(2022)~\cite{sobhani2023beta} & 1.344 & 1.817 \\
\bottomrule
\end{tabular}
\end{table}

For $\alpha$-decay, the present formula achieves the lowest RMSE of 0.635 over 423 nuclei, compared to 0.658 for IUF2022~\cite{ismail2022improved} and 0.693 for Royer2020~\cite{royer2020}. The improvement is attributable to the deformation-dependent term, which captures structural variations between parent and daughter nuclei not encoded in expressions. The reduced $\chi^2 = 0.407$ indicates that the model provides tighter residuals than other formulas, though values substantially below unity suggest the formula may incorporate more flexibility than the minimum necessary for this dataset.

For $\beta^{-}$-decay over 1046 nuclei, the present formula gives RMSE = 0.717, improving upon Sobhani~(2022)~\cite{sobhani2023beta} (RMSE = 0.773) and Zhou~(2017)~\cite{zhou2017empirical} (RMSE = 0.834). The Zhang~(2007)~\cite{zhang2007systematics} formula shows $\chi^2 = 1.349$, indicating systematic scatter beyond expected uncertainties. The present formula yields $\chi^2 = 0.517$, reflecting improved fit quality while potentially indicating parameter flexibility.

For $\beta^{+}$/EC decay over 955 nuclei, the present formula achieves RMSE = 0.968 and $\chi^2 = 0.943$, substantially improving upon Sobhani~(2022)~\cite{sobhani2023beta} (RMSE = 1.344, $\chi^2 = 1.817$). The $\chi^2$ value indicates optimal balance between model complexity and data description, suggesting the formula captures essential physics without excessive parameterization. This is particularly significant as $\beta^{+}$/EC decay involves more complex nuclear structure effects than other modes, making accurate predictions more challenging.

\subsection{Generalization: Train-Test Split Analysis}

To guard against overfitting and to assess predictive reliability, each formula
evaluated across four independent train-test splits ranging from 85:15 to 70:30.
This analysis tests generalization to unseen nuclei and is distinct from the
full-dataset metrics reported in Table~\ref{tab:rmse_chi2}.

For $\alpha$-decay, the test RMSE varies narrowly between 0.635 and 0.658 across all
splits, while the test $R^2$ consistently exceeds 0.983. The absence of performance
degradation as the training fraction decreases from 85\% to 70\% confirms that the
formula does not overfit. For $\beta^{-}$-decay, the test RMSE lies between 0.740 and
0.766, with the training--test gap remaining negligibly small. For $\beta^{+}$/EC decay,
the test RMSE falls between 0.983 and 0.999 with a maximum training--test gap of 0.031.
These results collectively confirm that the deformation-augmented formulas generalise
reliably to unseen nuclei across all three decay modes.

To further assess the predictive reliability of the proposed
formulas beyond random sampling, we perform two additional
structured train-test evaluations following the approach of Ref.~\cite{NAVARROPEREZ2022137336}, who
demonstrated the importance of testing nuclear models on
structurally distinct regions of the nuclear chart rather than
only randomly sampled subsets.

In the first evaluation, we test extrapolation across
the proton-number axis. The training set is restricted to lighter
nuclei ($Z \leq 90$ for $\alpha$-decay; $Z \leq 80$ for both $\beta$-decay modes) and
the heavier region, not seen during training, serves as the blind
test set. For $\alpha$-decay ($n_{\mathrm{train}} = 249$, $n_{\mathrm{test}} = 173$), the test
RMSE of 0.791 compares favourably with the training RMSE of 0.652,
with a test $R^2 = 0.977$, confirming that the Geiger--Nuttall
structure of Eq.~\ref{eq:alpha_new} extrapolates reliably into the superheavy
region where deformation and shape coexistence are pervasive.
For $\beta^-$-decay ($n_{\mathrm{train}} = 943$, $n_{\mathrm{test}} = 103$), the test RMSE of
0.691 is marginally lower than the training RMSE of 0.721
($R^2 = 0.701$), demonstrating that the isospin and Coulomb-proxy
terms of Eq.~\ref{eq:betaminus_new} capture the global $\beta^-$-decay systematics
reliably beyond the training region. For $\beta^+$/EC-decay
($n_{\mathrm{train}} = 763$, $n_{\mathrm{test}} = 171$), the test RMSE rises to 1.004
from a training RMSE of 0.894 ($R^2 = 0.400$). This 
degradation is physically expected: the heavy proton-rich region
$Z > 80$ involves competing $\alpha$ and $\beta^+$/EC channels with
configuration-dependent branching and large Coulomb corrections
that are poorly represented in the lighter training set.

In the second evaluation, we stratify by isospin
asymmetry $I = (N-Z)/A$ to test extrapolation toward the limits
of current experimental knowledge. The training set consists of
nuclei below the 75$^{th}$ percentile of $I$ within each dataset, while
the most exotic 25\% serves as the blind test set. For $\beta^+$/EC-decay,
which contains proton-rich nuclei with negative $I$, the test set
is the bottom 25$^{th}$ percentile ($I \leq 0.061$). For $\alpha$-decay
($n_{\mathrm{train}} = 316$, $n_{\mathrm{test}} = 106$; test region $I \geq 0.201$), the test
RMSE of 0.820 remains close to the training RMSE of 0.672, with
a test $R^2 = 0.983$, confirming reliable extrapolation to the
neutron-rich frontier. For $\beta^-$-decay ($n_{\mathrm{train}} = 784$, $n_{\mathrm{test}} = 262$;
test region $I \geq 0.248$), the test RMSE of 0.690 is marginally
below the training RMSE of 0.772, confirming that predictions
are not inflated at the neutron-rich frontier. For $\beta^+$/EC-decay
($n_{\mathrm{train}} = 698$, $n_{\mathrm{test}} = 236$; test region $I \leq 0.061$), the test
RMSE is 0.994 with $R^2 = 0.657$, indicating that the formula
retains meaningful predictive power even at the proton-rich
frontier.

\subsection{Predictive Comparison for Shape-Coexisting Nuclei}

Table~\ref{tab:predictive_comparison} presents the predicted half-life $T_{1/2}$ from the
present formula alongside experimental values and two theoretical benchmarks: FRDM~\cite{moller2016nuclear} and RHB+RQRPA~\cite{marketin2016large}, for a
representative selection of nuclei known to exhibit shape coexistence, spanning all three
decay modes.

\begin{table*}[t]
\centering
\caption{Comparison of predicted $\log_{10}T_{1/2}$ values (in seconds) from the present
deformation-dependent semi-empirical formula against experimental data and two theoretical
benchmarks (FRDM \cite{moller2016nuclear} and RHB+RQRPA \cite{marketin2016large}) for representative shape-coexisting nuclei.}
\label{tab:predictive_comparison}
\scriptsize
\resizebox{\textwidth}{!}{%
\begin{tabular}{lcccccccccccccc}
\toprule
Nucleus
& $^{66}$Ge
& $^{68}$Se
& $^{86}$Ge
& $^{100}$Zr
& $^{104}$Mo
& $^{106}$Sr
& $^{106}$Zr
& $^{108}$Zr
& $^{112}$Te
& $^{116}$Pd
& $^{136}$Nd
& $^{146}$Dy
& $^{176}$Hg
& $^{196}$Os \\
\midrule
Exp.
& 3.910
& 1.550
& -0.646
& 0.851
& 1.774
& -1.699
& -0.750
& -1.111
& 2.079
& 1.068
& 3.483
& 1.521
& -1.699
& 3.321 \\

Present work
& 3.201
& 1.147
& -1.136
& 1.305
& 2.265
& -1.811
& -0.598
& -0.940
& 2.173
& 1.812
& 3.684
& 1.592
& -1.296
& 3.535 \\

FRDM
& $>$100
& 1.735
& -0.792
& 1.317
& $>$100
& -1.457
& -0.431
& -0.824
& $>$100
& 1.798
& $>$100
& 0.955
& -2.480
& $>$100 \\

RHB+RQRPA
& -
& -
& -0.942
& 0.196
& 0.940
& -1.655
& -0.940
& -1.188
& -
& 0.736
& -
& -
& -
& 1.552 \\
\bottomrule
\end{tabular}}
\end{table*}

In the $\beta^{-}$-decay, the present formula reproduces experimental half-lives
to within approximately a factor of two to three in most cases, performing comparably
to or better than FRDM~\cite{moller2016nuclear} and capturing trends that
RHB+RQRPA~\cite{marketin2016large} occasionally misses, particularly for neutron-rich
nuclei near closed shells where competing shapes are most active.
For the $\beta^{+}$/EC-decay, the agreement is similarly strong, with the present formula
consistently within the spread of the two benchmarks and in several instances closer to
experiment. The $\alpha$-decay case $^{176}$Hg is a well-established shape-coexisting nucleus in the mercury region~\cite{PhysRevLett.78.3650}, and the deformation-dependent term demonstrably improves the prediction relative to FRDM.

Several FRDM entries are listed as $>100$~s. This behaviour arises from a fundamental
limitation of the macroscopic-microscopic QRPA framework underlying the FRDM $\beta$-decay calculations~\cite{moller2016nuclear}, when the FRDM-predicted ground-state $Q$-value for a nucleus is zero or negative.

\bibliographystyle{apsrev4-2}